\documentclass[conference]{IEEEtran}
\usepackage{cite}

\ifCLASSINFOpdf
\else
\fi
\ifCLASSOPTIONcompsoc
 \usepackage[caption=false,font=normalsize,labelfont=sf,textfont=sf]{subfig}
\else
 \usepackage[caption=false,font=footnotesize]{subfig}
\fi
\usepackage{graphicx}
\usepackage{pifont}

\begin{document}
%
\title{Secure, Robust, and Energy-Efficient Authenticated Data Sharing in UAV-Assisted 6G Networks}

\author{\IEEEauthorblockN{Atefeh Mohseni Ejiyeh}
\IEEEauthorblockA{University of California, Santa Barbara\\
atefeh@ucsb.edu}
}


%


\maketitle

\begin{abstract}
This paper confronts the pressing challenges of sixth-generation (6G) wireless communication networks by harnessing the unique capabilities of Unmanned Aerial Vehicles (UAVs). With the ambitious promises of 6G, including ultra-reliable 1 Tbps data delivery and ultra-low latency, the demand for innovative solutions becomes imperative. Traditional terrestrial base stations, though effective, exhibit limitations in scenarios requiring ubiquitous connectivity, prompting the integration of UAVs. In response to these challenges, we introduce a comprehensive solution. This involves UAVs collaboratively downloading desired content from service providers, and subsequently establishing secure connections with users for efficient content exchange. Accordingly, we introduce two new protocols: a collaborative group data downloading scheme among UAVs called SeGDS, and SeDDS for secure direct data sharing through out-of-band autonomous Device-to-Device (D2D) communication. Leveraging certificateless signcryption and certificateless multi-receiver encryption, these protocols offer lightweight, certificate-free solutions with features such as user revocation, non-repudiation, and mutual authentication. Prioritizing high availability, the proposed protocols effectively detect Denial of Service (DoS) and free riding attacks. A thorough evaluation underscores the superiority of the proposed protocols in both security and efficiency over existing models; SeDDS reduces overall computation by 3x, imposing a lighter communication load on UAVs, while SeGDS meets swarm UAV security requirements, reducing communication costs by 4x with low computation cost.

\end{abstract}


%
\IEEEpeerreviewmaketitle

\section{Introduction}
\label{sec:intro}

The sixth generation of wireless communication networks (6G) promises unparalleled advancements, featuring ultra-reliable 1 Tbps data delivery (100 times that of 5G), ultra-low latency below 100ms, and enhanced energy efficiency \cite{6groad}. These advancements open avenues for groundbreaking applications, from real-time immersive experiences to emergency networks. However, achieving these ambitious goals requires innovative infrastructure solutions.

\begin{figure}
  \centering
\includegraphics[width=0.9\columnwidth]{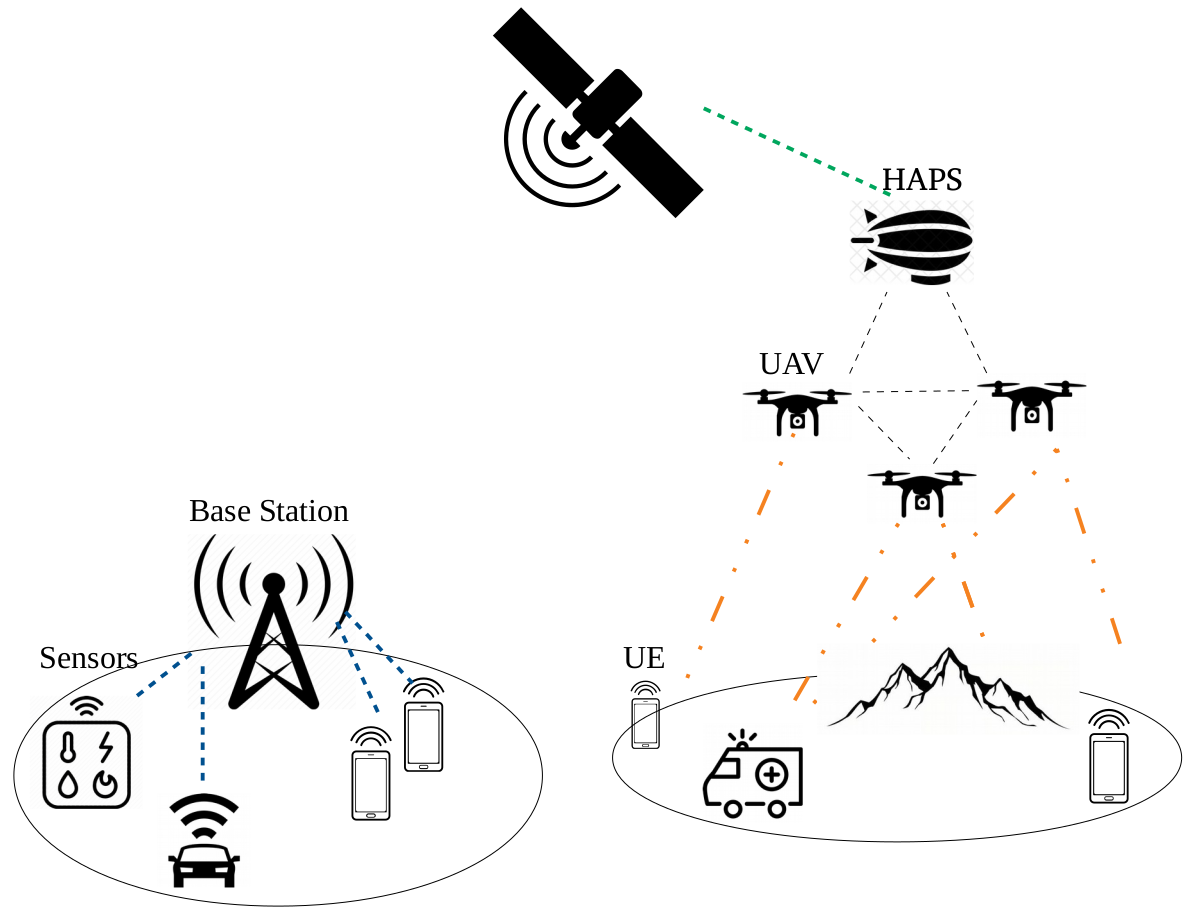}
\caption{The architecture of UAV-assisted networks in B5G/6G} 
\label{fig:arch}
\end{figure}

Traditionally, terrestrial base stations handle network coverage, facing limitations in scenarios demanding ubiquitous connectivity, such as disaster response or remote area communication. Unmanned aerial vehicles (UAVs), or drones, emerge as a solution \cite{swarmuav6g}. Their dynamic mobility and proximity, as depicted in Fig \ref{fig:arch}, allow them to fill coverage gaps and adjust network resources dynamically, addressing critical service delivery needs. UAVs can enhance communication reliability and availability through proactive caching for data dissemination \cite{uav-dist-filesh}, task offloading from resource-limited devices \cite{bai2023lowcost}, or providing emergency networking capability after natural disasters\cite{lin2021adaptive}.

While studies have explored networking paradigms for adopting UAV in the context of next generation of mobile networks \cite{emergencyUAV}, security concerns remain in their infancy \cite{dronesok}. Notably, recent studies identified vulnerabilities to localization attacks \cite{shootm-ahmad}, and lack of robust access control and authenticated message exchange among UAVs \cite{dronesecsur}.

Unlocking the capabilities of UAV-assisted 6G necessitates tackling a central challenge: the establishment of secure communication channels among entities\cite{shvetsov2023federated}. Primarily, as UAVs collaborate to download data for performance improvement and service delivery, security becomes a critical concern. Unsecured channels expose sensitive data to interception and manipulation, jeopardizing data privacy and network integrity. Additionally, ensuring reliable data delivery to end-users becomes imperative for minimizing delays and errors, especially in real-time applications and emergency networks.

This paper addresses both challenges by presenting two innovative protocols for secure and energy-efficient content sharing in UAV-assisted 6G networks. 
\begin{itemize}
    \item \textbf{Group Data Sharing Protocol} enables efficient content sharing among collaborating UAVs by downloading data packets, minimizing redundancy, and expediting delivery.
    
    \item \textbf{Direct Data Sharing Protocol} ensures secure and non-repudiable direct data sharing between UAVs and end-users, eliminating intermediate hops for tamper-proof transmission without third-party intervention.
\end{itemize}

By meticulously analyzing the security, performance, and overhead of these protocols against state-of-the-art standards, we showcase their superiority. This research establishes the groundwork for resilient and dependable communication channels in the evolving 6G era, unlocking the complete potential of UAV-assisted networking.
\section{Background}
\label{sec:solution}
\subsection{Extensible Authentication Protocol (EAP) in Beyond 5G} In the context of 5G and emerging 6G networks, Extensible Authentication Protocol with AKA (EAP-AKA) is pivotal for secure User Equipment (UE) authentication. This process centers around the Authentication Server Function (AUSF), a central entity managing user identities and cryptographic keys. EAP-AKA involves a cryptographic handshake initiated by the UE (or UAV), with the AUSF verifying the identity using a shared secret, establishing trust for encrypted communication. The AUSF's dual role includes issuing unique device IDs for verification in distributed protocols and facilitating high-level authentication for access to sensitive data \cite{ausf5g}. In our proposed protocols, UAVs use legitimate network IDs for registration in the data sharing schemes detailed in Sec \ref{sec:segd}.

\subsection{Device-to-Device (D2D) Communications}
D2D communication enables direct interaction between nearby devices, providing benefits like improved latency and personalized content delivery. In cellular-controlled D2D, the base station manages connections for optimal network performance. Unlicensed-spectrum-based D2D, utilizing Wi-Fi Direct or Bluetooth, offers autonomy and flexibility \cite{d2d6gsurvey}. D2D applications extend beyond mobile networks; Swarm UAVs group nearby UAVs for efficient content sharing \cite{d2dbeyond-v}-\cite{swarmuav6g}, and Cellular-Assisted V2X enhances direct communication between vehicles for improved safety and traffic management \cite{5gcontentshare-sign}. In this paper, we introduce a novel secure authentication and message exchange protocol for unlicensed-spectrum D2D, a first in the field. Additionally, we present a cluster-based D2D protocol operating on licensed spectrum, enhancing secure collaborative content sharing among UAVs in proximity.

\subsection{Certificateless Signcryption}
Certificateless signcryption (CL-signcryption) offers a cryptographic primitive combining digital signature and encryption functionalities into a single operation. Instead of relying on pre-issued certificates, users derive their keys based on their identities, simplifying key management compared to certificate-based schemes. CL-signcryption offers three main functionalities:

\begin{enumerate}
    \item \textbf{Signature}: It allows a user to prove ownership of their identity while authenticating data integrity and non-repudiation.
\item \textbf{Encryption}: It enables secure data confidentiality by encrypting information for specific intended recipients using their identities.
 \item \textbf{Signcryption}: It combines both functionalities, simultaneously signing and encrypting data for authenticity and confidentiality in a single step.
\end{enumerate}
In this paper, we employed \cite{clgsc} for its efficient certificateless signcryption functionalities.

\subsection{Certificateless Multireceiver Encryption}

Certificateless Multireceiver Encryption is a cryptographic approach devised to secure data transmission in multicast scenarios without the need for traditional public key certificates. It addresses scalability and efficiency challenges inherent in conventional multicast encryption by streamlining the key distribution process.

\begin{enumerate}
\item \textbf{Multireceiver Encryption (CL-MREnc)}: Sender utilizes the group leader's public key and unique identifiers for each recipient, then generates a multicast session key for encrypting the data, ensuring confidentiality.

\item \textbf{Multireceiver Decryption (CL-MRDec)}: Recipient employs their unique identifier and private key to derive the multicast session key.
\end{enumerate}

CL-MREnc streamlines key management, improves scalability, and offers an effective solution for secure multicast communication, notably advantageous in scenarios like UAV-assisted networks. Here, we leverage \cite{certificateless} for the proposed secure collaborative data sharing scheme detailed in Sec. \ref{sec:segd}.

\subsection{Threat Model} Adhering to the Dolev-Yao threat model \cite{dolev}, adversaries possess the capability to intercept messages transmitted through public channels, initiate and receive conversations with other participants, and engage in entity impersonation. We assume the Service Provider (Cloud) operates honestly and provides the original contents. Individual group leaders (UAVs) are considered rational entities with limited resources, behaving maliciously only when there's a perceived benefit. Collisions among adversaries are possible.

Resilience against stolen credential attacks is an open research question for UAVs \cite{ozmenshort}. Although, the propsoed collaborative scheme remains secure against stolen credentials of individual group members, implementing hardware security mechanisms to fortify both protocols against stolen or compromised credentials is planned for future work.
\section{SeGDS: Secure Group Data Sharing Scheme}
\label{sec:segd} 
In this section, the proposed cooperative data sharing protocol is explained step by step. It is assumed that the group members have selected a leader ($CH$) among themselves before starting the protocol.

\begin{table}[h]
\centering
\resizebox{\linewidth}{!}{%
\begin{tabular}{|c|l|}
\hline
\textbf{Term} & \textbf{Description}\\ \hline
\hline
$CH$, $SP$  & Coordinator Head, Service Provider \\ \hline
$FN$, $FS$ & File Name, File Segment\\ \hline
$CLGSC(ID_i,0,*)$ & Signature of entity with $ID_i$ \\ \hline
$CLGSC(ID_i,ID_j,*)$ & Signcryption of entity with $ID_i$ to $ID_j$ \\ \hline
\end{tabular}
}
\caption{Notation definitions}
\label{tab:comm-terms}
\end{table}

\subsection{ System Initialization Phase}
AUSF selects a cyclce group $G$ with generator $P$ on an elliptic curve with two prime numbers $p$, and $q$ given the security parameter $k$. Additionally, the AUSF selects $x_0$ as the private master key and $Q=x_0P$ as the public master key. The four hash functions are selected to be $H_0:\{0,1\}^*\times G\times G\times G \rightarrow Z_{q}^*$, $H_1: G \times \{0,1\}^* \times \{0,1\}^* \rightarrow Z_{q}^*$, $H_2: G \times G \rightarrow \{0,1\}^n$, and $H_3$ equivalent to SHA-256 and the utilized symmetric encryption function is AES-256.

\subsection{UAVs registration phase}
UAVs submit membership requests containing their identifier ($ID_{i}$), partial secret key $x_i \in Z_{q}^*$ for $i>=1$, and public partial key $X_i=x_iP$. The AUSF operation verifies these requests, ensuring access authorization and UAV authentication. Following successful verification, the AUSF encrypts the UAV's partial keys $(z_i, Y_i)$ as $z_i=y_i+x_0H_0(ID_i,Y_i, X_i, X_0)$ when $y_i \in Z_{q}^* , Y_i=y_iP$ and transmits them through a secure channel. The full private key of the entity with $ID_i$ is ($x_i,z_i$).

\subsection{Session Setup Phase}
The designated group leader establishes a secure session with the content service provider. It involves shareing the target file name $FN$ and generating a data encryption key ($K_d$) through Diffie–Hellman key exchange protocol. Consequently, the content server provider utilizes $K_d$ to encrypt all member requests pertaining to various segments of the requested data.

\subsection{Task Assignment and Cooperative Download Phase}
\label{gphase}
Group leader distributes downloading tasks among members equally or in varied sized segments given the total file size and UAVs QoS.

\textbf{Step 1}: Group leader divides the target file segments among members. Messages sent to UAVs, e.g., UAV with IDi is $M_1:(ID_{ch},ID_i, FN, FS, T_s, CLGSC(CH,0,*))$
indicate a file segment $FS$ to download. The timestamp $T_s$ marks task assignment, and the group leader's signature on the message is represented by $CLGSC(CH,0,*)$.

\textbf{Step 2}: $UAV_i$ verifies the group leader's signature upon receiving the message. Upon confirmation, the UAV forwards the message to the content service provider to download their portion of the file.

\textbf{Step 3}: The content service provider, upon receiving the UAV's request, verifies the freshness of the timestamp and the validity of the group leader's signature. Upon confirmation, it encrypts the relevant segment of the file data associated with $FN$ with $K_d$ and delivers it to the $UAV_i$ as $M_2:(ID_{sp},ID_i,FN,FS,C_i,T_s,\sigma_{sp}=CLGSC(SP,0,H_3(C_i)||*)$ where $C_i=Enc(K_d, M_i)$.

\subsection{Data Sharing Phase}
In this phase, each UAV shares their downloaded segment with others, verifying and accepting received messages. After completing the download of $FS$ by the UAV, they verify the content service provider's signature. If confirmed, they append the download completion timestamp ($T_d$) and broadcast it among all members as $M_3=(M_2, T_d, \sigma_i=CLGSC(ID_i,0,(M_2||T_d))$. Each member verifies received messages and accepts the segment data upon confirmation. The group leader also involves in downloading a file segment, verifies received messages, and maintains a blacklist for UAVs shared invalid messages. If a UAV fails to perform their assigned task within a specified period, the leader assumes non-cooperation and adds their ID to the blacklist. If any data segments remains, the group leader repeats \ref{gphase} steps.

\subsection {Data Consolidation Phase}
After having all the data segments downloaded, the group leader encrypts the file encryption key $K_d$ using a multi-receiver encryption scheme and shares it among members as $M_4: CL-MREnc(K_d),T_f, GLGSC(CH,0,*)$. A UAV added to the group leader's blacklist should not be able to decrypt the value of the key. Each member, upon receiving the message, verifies the signature, checks the timestamp freshness $T_f$, and decrypts the data using CL-MRDec with the obtained data key through $M_i=Dec(K_d, C_i)$.

\section{SeDDS: Secure Direct Data Sharing scheme}
\label{sec:sedds} 
To directly share data between UAVs and user equipment (UE) the proposed SeDDS protocol is described here.

\textbf{Step 1}: $UE_i$ sends its request to $UAV_j$ through $(ID_i || ID_j || FN || T_s || g^a || GLGSC(ID_i,0,*))$ which $FN$ describes desired file name.

\textbf{Step 2}: $UAV_j$ verifies the validity of the received request; then chooses a random $b \in Z_q^*$, computes $k_c =(g^a)^b$ $UAV_j$, encrypts the content $M$ associated with $FN$ through $M'=Enc(k_c,M)$, and transmits the cipher data as $(ID_j || ID_i || M' ||T_s ||\sigma_{s p}|| \sigma_j=GLGSC(ID_j,0,*)))$ to $UE_i$.

\textbf{Step 3}: By receiving the message of $UAV_j$, $UE_i$ verifies the $\sigma_j$: if it holds true, $UE_i$ sends a key hint request to the $UAV_j$ containing $(ID_i || ID_j || FN || T_s || T_i || FN \oplus h(M') || \sigma_{i,1}=CLGSC(ID_i,0,*)$.

\textbf{Step 4}: Upon receiving a key hint request, $UAV_j$ sends the data encryption key to $UE_i$. To mitigate payment risk and enhance user experience, a prepayment strategy is employed. $UE_i$ is subject to pay a portion of the credits ($0<p<100$) after receieving the encrypted data in step 3. Only if $UE_i$ verifies the data's originality, $UAV_j$ informs the service provider about the full credit charges. $UAV_j$ sends a message $(ID_j||ID_i||g^b||FN||\sigma_j=CLGSC(ID_j,0,*))$.

\textbf{Step 5}: $UE_i$ calculates the data encryption key using $(g^b)^a$ and decrypts $M'$. It verifies the signature of the service provider, $\sigma_{sp}$. If the signature is valid, $UE_i$ sends a message $(ID_i||ID_j||FN||FN\oplus h(M)||T_i||\sigma_{i,2}=CLGSC(ID_i, 0, *))$.

$UAV_j$ can forward the Ack message from $UE_i$ to the service provider. AUSF checks the validity of $UE_i$'s signatures. If verified, it computes a hash of the original content ($FN$) through $FN \oplus h(M'')$. If $UAV_j$ only received $\sigma_{i,1}$ and the XOR result doesn't match, it indicates a failed connection. On the other hand, if $UAV_i$ received $\sigma_{i,2}$, it signals a successful data transfer. The protocol can also be applied for data exchange between two UEs, reducing the load on UAVs. AUSF may store anonymized records on user content possessions for motivation, proximity-based services and reward users for active participation.
\section{Evaluation}
\label{sec:evaluation}
\begin{figure*}
\centering
\begin{tabular}{cccc}
    \subfloat[Direct Data Sharing Performance Comparison]{\includegraphics[width=0.33\textwidth]{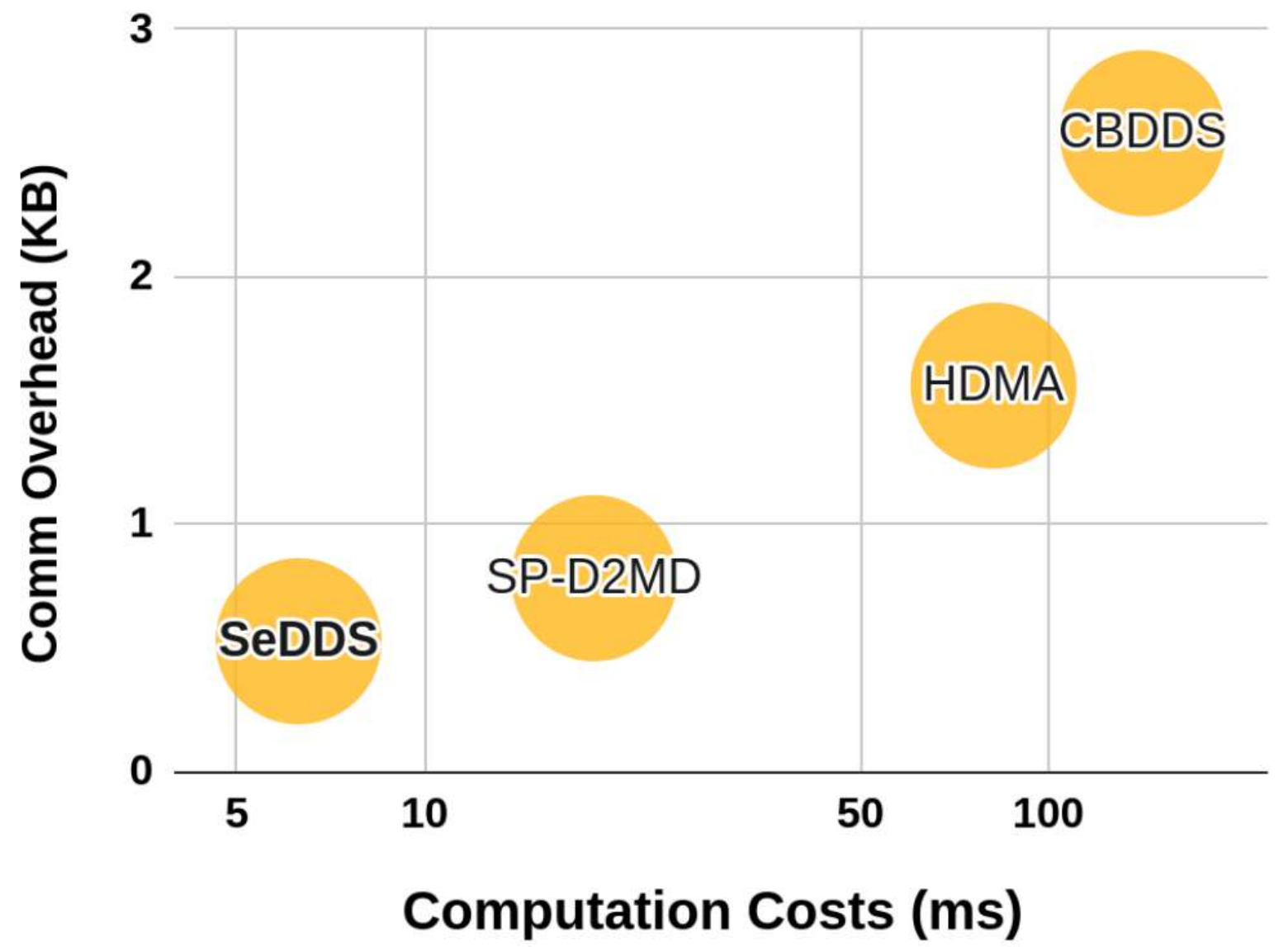}} &
    \subfloat[Group Communication Overhead Comparison]{\includegraphics[width=0.33\textwidth]{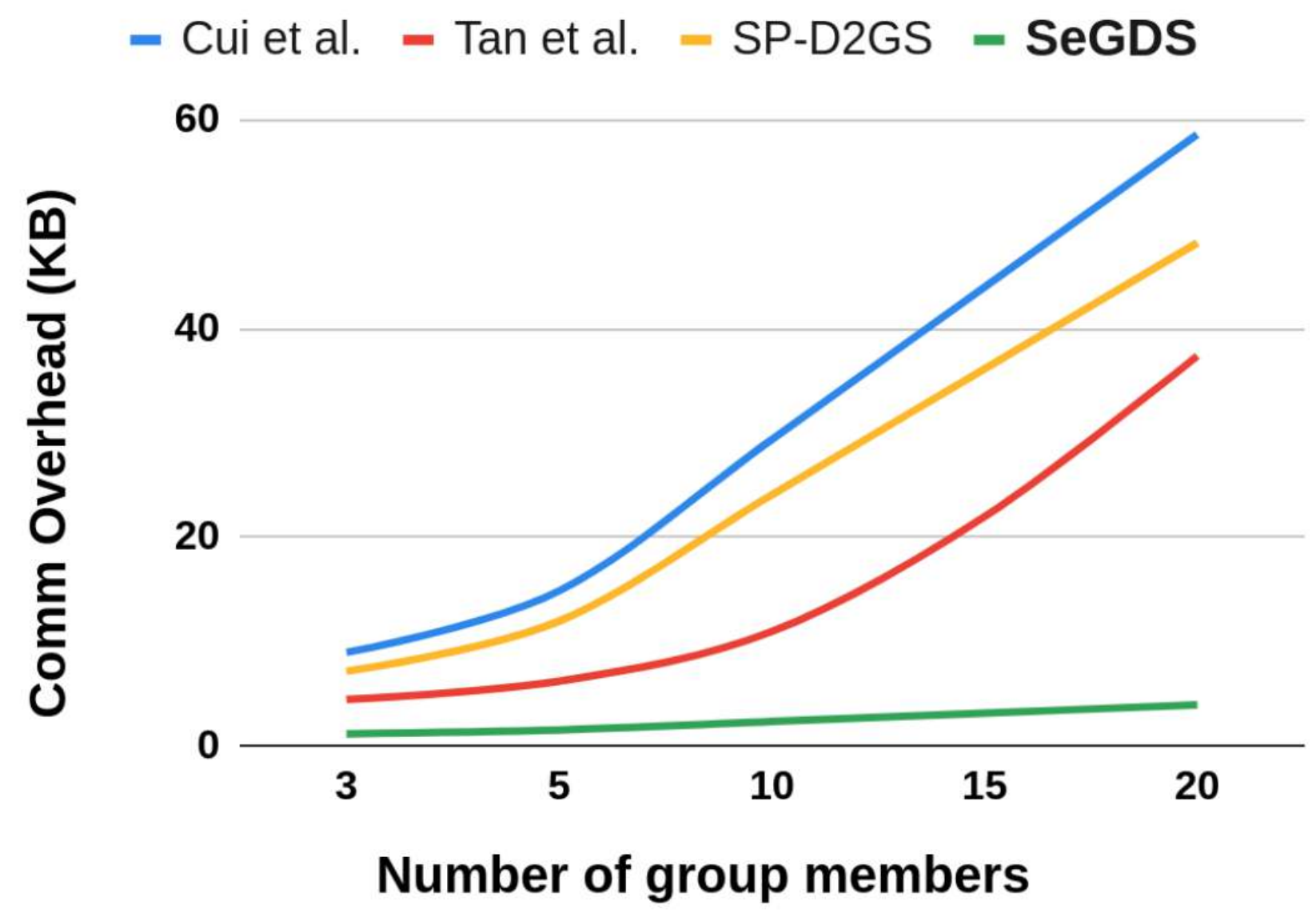}} & \subfloat[Group Computation Overhead Comparison]{\includegraphics[width=0.33\textwidth]{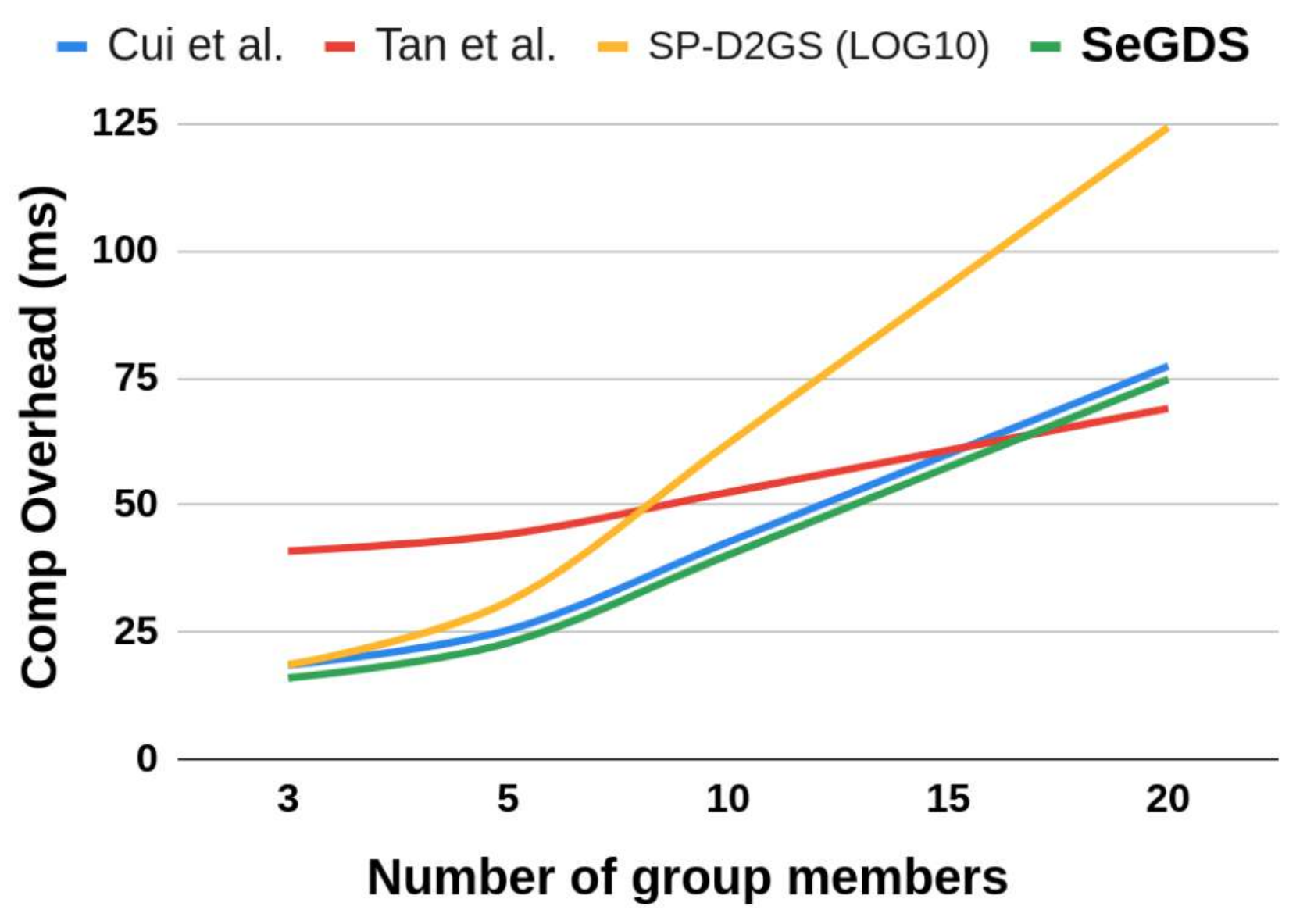}} & \\
\end{tabular}
\caption{Performance Evaluation of Proposed Schemes: SeDDS Fig (a), and SeGDS Fig (b,c) vs. State of the Art}
\label{fig:all}
\end{figure*}

In this section, we compare the security features, computational and communication overhead of our proposed schemes with other related work. Due to limited literature on secure data sharing in UAV communication \cite{dronesecsur}; we compare our proposed protocols with the related scheme in similar networks such as VANET as depicted in Fig \ref{fig:all}.

\subsection{Performance Analysis}
In the subsequent sections, we quantify the computation and communication overhead introduced by our protocols and compare them with state-of-the-art methods. The denoted terms include execution time for: $T_{e}$ modular exponential operation (1024 bits), $T_{m}$ scalar multiplication in cycle group G related to ECC, $T_{sm}$ scale multiplication in a bilinear group, $T_{bp}$ bilinear pairing, $T_{AES}$ AES-256 encryption or decryption, $T_{RSA}$ RSA signature generation, encryption, verification, or decryption, $T_{s}$ digital signature using the ECDS algorithm, and $T_{pa}$ for point addition operation. Negligible operations, such as hash functions, have been omitted for simplicity. Tables \ref{tab:dir-cost} and \ref{tab:coop-cost} delineate the communication and computation costs, respectively, in direct and collaborative communication modes utilizing a Raspberry Pi 3 model B.

HDMA \cite{hdma} is a 5G-enabled VANET hybrid message authentication scheme. It involves vehicles authenticating with Road Side Boxes to obtain local group signature private keys, using them to sign messages directly to other vehicles. To enable a fair comparison with our work, we incorporate a 3-step data exchange post-authentication phase into their overall overhead. Despite meeting specific security requirements, HDMA is less suitable for UAV scenarios due to its reliance on costly public cryptography and the absence of non-repudiation, a crucial security element for UAVs \cite{swarmuav6g}.

CBDDS \cite{cbdds-dist-cach} introduces a secure and revocable cache-based distributed data sharing scheme for privacy-preserving authentication between vehicles and Edge Servers (ES) in vehicular networks. In their approach, ES takes on roles of authentication and authorization, incorporating a token authentication mechanism and a multi-authority CP-ABE\footnote{Cipher Policy Attribute Based Encryption} algorithm for access control. However, their assumption of ES lacking resource limitations is not suitable for UAVs in our system model, and the computation overhead grows with the number of supported attributes.

The authors of \cite{drone-2p-military} focused on UAV communication security in military zones, proposing two protocols: SP-D2GCS for communication between a drone and Ground Control Station, and SP-D2MD for communication between a drone and a monitoring drone, acting as a middleman to the ground station. Despite meeting some security requirements, their model relies on the monitoring drone transmitting all data, unlike our approach where the group leader downloads its own segment, similar to other UAVs. As a result, their protocols are not suitable for (1) resource-limited drones and (2) emergency networks when a connection to the ground station is unavailable.

In conclusion, SeDDS not only surpasses the state of the art in security and scalability but, as demonstrated in Table \ref{tab:dir-cost}, reduces overall computation by 3x and imposes a lighter communication load on UAVs.

\begin{table}[h]
\centering
\resizebox{\linewidth}{!}{%
\begin{tabular}{|c|l|c|}
\hline
\textbf{Scheme} & \textbf{Computation cost (ms)} & \textbf{Tot bytes}\\ \hline
HDMA\cite{hdma} & $26T_e+5T_{RSA}+2T_{AES}$ = 81.66 & 1560 \\ \hline
CBDDS\cite{cbdds-dist-cach} & $15T_e+13T_{pb}+9T_{sm}+4T_{RSA}+T_{AES}$= 141.65 & 2580 \\ \hline
SP-D2MD\cite{drone-2p-military} & $2T_e+4T_{m}+4T_{AES}$ = 18.7 & 781 \\ \hline
\hline
\textbf{SeDDS} &$2T_e+10T_m+2T_{AES}$ = \textbf{6.28} & 526 \\ \hline
\end{tabular}
}
\caption{Comparative computation and communication costs analysis for Secure Direct Data Sharing Schemes}
\label{tab:dir-cost}
\end{table}

Cui et al. \cite{coop-download-v} introduced a secure cooperative data downloading scheme leveraging edge computing in VANETS. The method relies on RSUs\footnote{Road Side Units} caching popular data in nearby edge computing vehicles (ECVs), facilitating direct downloads by vehicles. However, the assumption of sufficient available qualified ECVs without resource constraints and the high communication overhead making it impractical for dynamic and UAV-assisted emergency networks.

Tan et al. \cite{uav-v2v-grp} present a certificateless mutual authentication protocol for UAV group association, employing a shared partial secret key for secure UAV group key generation, updates, and dynamic revocation. They suggest a method for remote V2V message dissemination through decentralized edge RSUs, relying on pre-stored records. However, the scheme's dependence on vehicular cloud authentication and the assumption of an existing wired RSU cluster constrain its practicality, particularly for emergency networks or out of band direct connections between users and UAVs.

Therefore, SeGDS, the proposed group data sharing scheme, not only meet the swarm UAVs security requirement but also it poses less computation costs and reduces the total communication cost by 4x as as depicted in Table \ref{tab:coop-cost}.

\begin{table}[h]
\centering
\resizebox{\linewidth}{!}{%
\begin{tabular}{|c|l|c|}
\hline
\textbf{Scheme} & \textbf{Computation cost (ms)}  & \textbf{Tot bytes}\\ \hline
Cui et al. \cite{coop-download-v} & $13T_m+3T_e+6NT_m$ = 25.51 & 14.9K \\ \hline
Tan et al. \cite{uav-v2v-grp}  & $32T_m+2T_{bp}+18T_e+2N(T_m+T_e)$=44.31 & 6.25K \\ \hline
SP-D2GS\cite{drone-2p-military}& $N(7T_{AES}+16T_m+12T_e)$ = 310.83 & 12K \\ \hline
\hline
\textbf{SeGDS} & $3(2N+3)T_m+2T_e$ = \textbf{22.72} & \textbf{1.5K} \\ \hline
\end{tabular}
}
\caption{Comparative computation and communication overhead in group data sharing Schemes (N members=5)}
\label{tab:coop-cost}
\end{table}

\subsection{Security Analysis}
This section conducts an in-depth security analysis of SeDDS and SeGDS protocols, with a comparison to existing work detailed in Table \ref{tab:sec-comparison}.

\begin{itemize}

\item \textbf{Confidentiality and Integrity:} To safeguard exchanged data from unauthorized access, SeGDS encrypts file content with AES, and even if an adversary acquires a UAV's private key, the group leader can revoke victim's credential, preventing access to the session key shared via multicast encryption. In SeDDS, UAVs generate a secure symmetric encryption key through DHKE based on the DLP\footnote{Discrete Logarithm Problem} assumption. Our protocols utilizes the certificateless signcryption \cite{clgsc} on all exchanged messages, ensuring confidentiality, data integrity, and unforgeability under the CDHP\footnote{Computational Diffie-Hellman Problem} and DLP assumptions.

\item \textbf{High Availability} ensures system resilience to disruptions like DoS attacks. Our protocols guarantee service availability through UAV-friendly computation constraints, identifying and excluding entities causing delays, and enforced authentication on all message exchanges. 

\item \textbf{Mutual Authentication} The Authentication Server Function (AUSF) plays a vital role in the 5G Core network, ensuring authentication for network entities among other responsibilities. This ensures the security of identity handling in our protocols. Additionally, mutual authentication occurs among all entities through signcryption of messages using CLGSC\cite{clgsc}. The scheme is resistant to forgery based on the DLP assumption. In both protocols, entities are required to verify the sender's signature at each step before advancing further.

\item \textbf{Non-repudiation} In UAV aided data sharing, it's crucial to ensure actions can't be denied\cite{swarmuav6g}. Signatures from the content service provider and UAVs prevent forgery and denial attacks, ensuring detectability and non-repudiation. The ID and timestamp on each message reveal participation details. In SeGDS, any attempt to share tampered data is easily detected. In SeDDS, malicious $UAV_j$ sharing tampered data with $UE_i$ requires a valid Ack message from $UE_i$ to prove an honest data exchange. This Ack message contains $P_i\oplus(M'')$ since AUSF uses the original data M from CSP to verify the signature. Unless $UAV_j$ forges $UE_i$'s signature, it can't deny malicious behavior. $UE_i$ cannot create a false report against $UAV_j$ as it requires forging the $UAV_j$ signature.

\begin{table*}[h]
\centering
\resizebox{\linewidth}{!}{
\begin{tabular}{|l|c|c|c|c|c|c|c|c|}
\hline
\textbf{Security Requirements} & HDMA\cite{hdma} & CBDDS \cite{cbdds-dist-cach} & SP-D2MD\cite{drone-2p-military} & Cui et al. \cite{coop-download-v} & Tan et al. \cite{uav-v2v-grp} & SP-D2GS\cite{drone-2p-military} & \textbf{SeDDS} & \textbf{SeGDS} \\ \hline

Confidentiality and Integrity &  \ding{51} &  \ding{51} &  \ding{51} &  \ding{51} &  \ding{51} &  \ding{51} &  \ding{51} &  \ding{51}\\ \hline

High Availability &  \ding{55} &  \ding{55} &   \ding{51} &   \ding{55} &   \ding{55} &   \ding{51} &  \ding{51} &  \ding{51} \\ \hline

Mutual Authentication &  \ding{51} &  \ding{51} &  \ding{51} &  \ding{51} &  \ding{51} &  \ding{51} &  \ding{51} &  \ding{51}\\ \hline

Non-Repudiation & \ding{55} &  \ding{51} &  \ding{55} &  \ding{51} &  \ding{51} &  \ding{51} &  \ding{51} &  \ding{51} \\ \hline

Content-agnostic &  \ding{51} &  \ding{55} &  \ding{51} &  \ding{55} &   \ding{55} &  \ding{51} &  \ding{51} &  \ding{51} \\ \hline

Support Offline D2D &   \ding{55} &   \ding{55} &   \ding{55} &  \ding{55} &  \ding{55} &  \ding{55} &  \ding{51} &   \ding{55}  \\ \hline

Support Group Data Sharing &   \ding{55} &   \ding{55} &   \ding{55} &  \ding{51} &  \ding{51} &  \ding{51} &  \ding{55} &   \ding{51}  \\ \hline

Support User/UAV Revocation&  \ding{55} &   \ding{51} &   \ding{55} &   \ding{55} &  \ding{51} &  \ding{55}  &   \ding{51} &   \ding{51}  \\ \hline

Resist Collusion attack &  \ding{51} &  \ding{51} &   \ding{55} &   \ding{55} &  \ding{51} &   \ding{55} &  \ding{51} &  \ding{51}  \\ \hline

Resist Free-ridding attack &  \ding{55} &   \ding{55} &   \ding{55} &  \ding{55} &  \ding{55} &   \ding{55} &  \ding{51} &   \ding{51}   \\ \hline
\end{tabular}
}
\caption{Security Comparison in Direct and Group Data Sharing Schemes.}

\label{tab:sec-comparison}
\end{table*}

\item \textbf{Content-agnostic} In UAV-assisted networks, the protocol's adaptability to diverse data types is crucial for applications like emergency networks \cite{emergencyUAV} or base station traffic offloading \cite{bai2023lowcost}. Unlike some approaches relying on contextual data, our protocols stay practical for real-world scenarios. Data selection in our protocols is directed by the group leader or UE, eliminating the dependency on historical data or static distributed storage systems.

\item \textbf{Support Offline Device-to-device (D2D) communication} in unlicensed spectrum enhances energy efficiency, boosts data rates, reduces link latency, and allows devices to utilize interfaces like Bluetooth or Wi-Fi Direct \cite{d2d6gsurvey}. Given the critical applications relying on this communication, such as public safety, its security is paramount, and, to the best of our knowledge, although there's D2D protocols with intervention of base station \cite{hdma}-\cite{cbdds-dist-cach}, there lacks a secure outbound autonomous D2D authentication and message exchange protocol except the proposed SeDDS scheme.

\item \textbf{Support Group Data Sharing} SeGDS achieves lightweight overhead as group size increases, ensuring efficient collaboration for tasks. Fig \ref{fig:all} (b,c) illustrates this effectiveness, maintaining support for mutual authentication, message integrity, and non-repudiation through strategic task splitting, preventing the need for full content sharing between the service provider and UAVs.

\item \textbf{Support user/UAV revocation} We utilize certificateless public key cryptography (CLPKC) with blacklist-based revocation, where user's private key stems from contributions by both the KGC (Key Generator Center, here AUSF) and the entity. The KGC, while unaware of the user's private key, can authenticate its public key. In proposed protocols, if the malicious behavior of a UAV or UE exceeds a predefined threshold during protocol execution, its ID is blacklisted, preventing further service use. Notably, CLPKC avoids the certificate revocation overhead on devices.

\item \textbf{Resist Collusion attack} A collusion attack involves authorized but malicious UAVs collaborating to provide malicious services to authorized entities. In SeGDS, there are two scenarios of collusion attacks. (1) A group of malicious UAVs colluding with a legitimate group leader UAV to obtain session keys of other groups without participating in the protocol. This is prevented as the data encryption key is unique per group session and generated by the group leader and service provider through a Diffie-Hellman protocol. (2) A group of malicious UAVs colluding to share tampered data with benign UAVs is detected, as the service provider signs every file segment with a proper timestamp before sharing with UAVs.

\item \textbf{Resist Free-riding Attack} Attempting to receive data without participating or paying characterize free-riding attacks. In SeGDS, detection of non-participating UAVs occurs during the data consolidation phase by the group leader. This detection prevents the decryption of the session key, given the exclusion from the multicast encryption scheme. In SeDDS, a $UE_i$ trying to receive data without acknowledging the recipient of the service faces two choices. They can either refrain from sending the key hint request (step 3), rendering the encrypted data unusable. Alternatively, they may attempt to create a fake report against $UAV_j$ by forging $UAV_j$'s signature, an impractical task under the DLP assumption.
\end{itemize}

\section{Related Work}\label{sec:related}
While prior research explores UAV communication using various networking paradigm and blockchain consensus \cite{luo2024escm}-\cite{wang2023blockchain}, our paper emphasizes securing data communication in UAV-assisted networks through authentication, key agreement, and data exchange protocols.

\subsection{Authentication and Key Agreement}
\cite{tan2023secure} proposes a secure and efficient authenticated key management scheme for internet of vehicles. Notably, their certificateless group authentication reduces public key overhead on devices. Addressing UAV-assisted VANETs, \cite{cui2023aka} introduces a lightweight and two-factor AKA scheme based on chaotic maps. To optimize storage of honeywords due to limited computational storage of vehicles and UAVs, a hybrid generation algorithm is proposed. Additionally, an intelligent drone-assisted anonymous AKA for 5G VANETs is designed in \cite{zhang2020intelligent}, utilizing the widely-used Real-Or-Random (ROR) model with reasonable computational overhead on UAVs.

\subsection{Secure Data sharing}
Considering the limited focus on secure data sharing in UAVs, insights are drawn from related work in adjacent networks, including D2D communications and VANET. In \cite{mohseni2018incentive}, authors propose a secure data sharing scheme between mobile devices in 5G, relying on base station to establish connections, with data flow occurring directly between devices. They introduce an incentive mechanism to encourage participation, tracking malicious behavior to ensure service availability and mitigate free-riding attacks. In the realm of 5G-enabled vehicular networks, a secure content-sharing approach presented in \cite{5gcontentshare-sign}. However, its dependence on a Trusted Authority (TA) to store buffers for content tracking limits its practicality for dynamic content sharing and emergency scenarios, as it requires TA availability for vehicle selection. A fast authentication and data distribution protocol in 3GPP 5G network presented in \cite{cao2018fast}, where a group of narrowband Internet of Things terminals achieves concurrent access authentication and data transmission. Addressing the need for a secure video reporting scheme in \cite{zhong2023secure} outlines a scheme where edge nodes perform message verification and classification to expedite report analysis and resolve duplicates.
\section{Conclusion}
In conclusion, this paper addresses the pressing demands for scalability, high availability, and security in wireless networks through the introduction of two protocols, SeDDS and SeGDS. SeDDS enables secure direct data sharing among devices in an out-of-band autonomous D2D communication setting, eliminating the need for a trusted party or presuming trust among entities. On the other hand, SeGDS presents a collaborative data sharing scheme, meeting security requirements for swarm UAVs while minimizing computation and communication costs. The meticulous evaluation of these protocols, demonstrated through comparative analyses, underscores their superiority in both security and efficiency over existing models. As we navigate the evolving landscape of 6G, these protocols provide a transformative pathway toward unlocking the full potential of UAV-assisted communication. Future work will focus on enhancing resistance against a broader range of attacks, particularly those targeting UAV hardware.
\label{lastpage}

\bibliographystyle{IEEEtran}
%
\bibliography{bare_conf}

\end{document}